\documentclass[aps,pra,twocolumn,showpacs,prabib, superscriptaddress]{revtex4}

\usepackage{graphicx}
\usepackage{amssymb,amsmath}
\usepackage{url}

\newcommand{\bea}{\begin{eqnarray}}
\newcommand{\eea}{\end{eqnarray}}
\newcommand{\be}{\begin{equation}}
\newcommand{\ee}{\end{equation}}
\newcommand{\rr}{\mathbf{r}}
\newcommand{\rhob}{\mbox{\boldmath$\rho$}}
\newcommand{\Omegab}{\mbox{\boldmath$\Omega$}}
\newcommand{\Omegabs}{\mbox{\scriptsize\boldmath$\Omega$}}
\DeclareMathOperator\atan{atan}
\DeclareMathOperator\argth{argth}

\DeclareMathOperator\argch{argch}
\DeclareMathOperator\ch{ch}
\DeclareMathOperator\sh{sh}
\DeclareMathOperator\re{Re}
\DeclareMathOperator\im{Im}

\begin{document}

\title{Third virial coefficient of the unitary Bose gas}
\author{Yvan Castin} 
\affiliation{Laboratoire Kastler Brossel, \'Ecole Normale Sup\'erieure, CNRS and UPMC, 24 rue Lhomond, 75231 Paris, France}
\author{F\'elix Werner}  
\affiliation{Laboratoire Kastler Brossel, \'Ecole Normale Sup\'erieure, CNRS and UPMC, 24 rue Lhomond, 75231 Paris, France}

\begin{abstract}
By unitary Bose gas we mean a system composed of spinless bosons 
with $s$-wave interaction 
of infinite scattering length and almost negligible (real or effective)
range. Experiments are currently trying to realize it with cold atoms. 
From the analytic solution of the three-body problem in a harmonic potential,
and using methods previously developed for fermions, we determine the third  cumulant (or cluster integral) 
$b_3$ and the third
virial coefficient $a_3$ of this gas, in the spatially homogeneous case, as a function of its temperature and
the three-body parameter $R_t$ characterizing the Efimov effect. A key point
is that, converting series into integrals (by an inverse residue method),
and using an unexpected small parameter (the three-boson mass angle
$\nu=\pi/6$), one can push the full analytical estimate of $b_3$  and $a_3$
up to an error that is in practice negligible.
\end{abstract}

\maketitle

\section{Introduction}

The field of quantum gases has been exploring, in the last decade, the strongly interacting regime, thanks to the possibility
of tuning the $s$-wave scattering length $a$ to arbitrarily large values (in absolute value) with the Feshbach resonance
technique \cite{Heinzen,Julienne}. This opens up the perspective of studying a fascinating object, the unitary gas, such that
the interactions among particles have an infinite $s$-wave scattering length and a negligible range.
The two-body scattering amplitude then reaches the maximal modulus allowed by unitarity of the $S$ matrix, and the gas is
maximally interacting.

For the spin-$1/2$ Fermi gases, the experimental realisation and characterisation of the strongly interacting regime have
been fully successful \cite{prems,BookZwerger}, recently culminating with the measurement of the equation of state
of the unitary gas, both at high and low temperature $T$  \cite{Nascimbene,Navon,Zwierlein}.
In the unpolarized case, this has allowed a precise comparison with the theoretical predictions, that are pushed to their limits.
At zero temperature, in practice $T/T_F \ll 1$ where $T_F$ is the Fermi temperature, the measurements have confirmed
the precision of the most recent variational fixed-node calculations, as far as the universal number
$\xi=\mu/(k_B T_F)$ is concerned, $\mu$ being  the chemical potential of the gas \cite{Gezerlis}. For the superfluid phase transition,
the experiments confirm the expected universality class and find a value of the critical temperature $T_c$ that
slightly corrects the result of the first Quantum Monte Carlo calculations \cite{Svistunov} and that confirms the one 
of most recent Quantum Monte Carlo calculations \cite{Goulko}. Above $T_c$, the measurements at MIT are in 
remarkable agreement with the diagrammatic Monte Carlo method \cite{BDMC}. Finally, in the non-degenerate regime
$T>T_F$, the experiments at ENS have been able to confirm the value of the third virial coefficient $a_3$ of the
spatially homogeneous gas, already theoretically deduced \cite{Drummond} from the analytical solution of the three-body
problem in a harmonic trap \cite{Werner} and reproduced later on by a diagrammatic method \cite{Leyronas}; these
experiments are even ahead of theory in getting the value of $a_4$, not yet extracted by theory in a reliable way from
the four-body problem  \cite{Blume}.

For the strongly interacting gases of spinless bosons, the experimental studies are less advanced, due to the Efimov effect \cite{Efimov}:
The effective three-body attraction predicted by Efimov, and leading to his famous weakly bound trimers, for which there
is now an experimental signature \cite{Revue}, leads to a strong  increase of atomic losses in the gas, due to three-body collisions
with strongly exothermic formation of deeply bound dimers. In the unitary limit,
one can at the moment prepare a stable and thermal equilibrium Bose gas only in the non-degenerate regime $\rho\lambda^3 \ll 1$ \cite{SalomonPetrov},
where $\rho$ is the gas density and $\lambda=(2\pi\hbar^2/m k_B T)^{1/2}$ is the thermal de Broglie wavelength:
The two-body elastic collision rate, scaling as $\frac{\hbar}{m} \rho \lambda$, then overcomes the three-body loss rate
scaling as $\frac{\hbar}{m} \rho \lambda^4$
\cite{Esry}\footnote{This holds within a dimensionless factor which is a periodic function of $\ln(R_t/\lambda)$~\cite{Werner_gen2,SalomonPetrov},
where $R_t$ is the three-body parameter.}.
Fortunately, there exists a few ideas to explore to reduce losses, such as taking advantage of the loss-induced Zeno effect in an
optical lattice \cite{Zoller}, or simply the use of narrow Feshbach resonances \cite{Petrov,Wang}.

On a theoretical point of view, the study of the unitary Bose gas is just starting. Most of the works do not take into account
in an exact way the three-or-more-body correlations \cite{Pandha,Stoof,Ho2012}; they cannot thus quantitatively account for
the fact that resonant interactions among bosons involve the three-body parameter  $R_t$, a length giving the global energy
scale in the Efimov trimer spectrum (no energy scale can be given by the scattering length here, since it is infinite). 
As a consequence, the various phases under which the unitary Bose system may exist at thermal equilibrium, as functions
of temperature, remain to be explored. At zero temperature, inclusion of a hard-core three-body interaction, 
allowing one to adjust the value of $R_t$ and to avoid the system collapse, has allowed one to show, with numerical
calculations limited to about ten particles, that the bosons form a $N$-body bound state, with an energy that seems
to vary linearly with $N$ \cite{vonStecher}, which suggests a phase of bounded density at large $N$, for example
a liquid. At high temperature (more precisely at low density $\rho\lambda^3\to 0$), the natural theoretical approach
is the virial expansion \cite{Beth}, that we recall here for the spatially homogeneous gas \cite{Huang}:
\be
\frac{P\lambda^3}{k_B T} = \sum_{n\geq 1} a_n (\rho \lambda^3)^{n}
\ee
where $P$ is the gas pressure. The coefficient $a_n$ is precisely the $n^{\mathrm{th}}$ virial coefficient.
In practice, it will be more convenient to determine the coefficients $b_n$ of the expansion of the grand potential 
$\Omega$ in powers of the fugacity $z=e^{\beta \mu}$, with the chemical potential $\mu$ tending to $-\infty$:
\be
\Omega = -\frac{V}{\lambda^3} k_B T \sum_{n\geq 1} b_n z^n,
\ee
with $V$ the system volume and $\beta=1/(k_B T)$. The coefficients $b_n$ are called cluster integrals in reference
\cite{Huang}.
Since $-\beta \Omega$ is the logarithm of the grand canonical partition function, it is natural to call
$b_n$ the $n^{\mathrm{th}}$ cumulant, as we shall do in this paper. The knowledge of all cumulants
up to order $n$ included allows one to determine all the virial coefficients up to the same order.
Note that $a_1=b_1=1$ by construction; the higher order expressions that are useful here are deduced in a simple
way from the thermodynamic equalities $\Omega=-PV$ and $N=-\partial_\mu \Omega$~\cite{Huang}:
\bea
\label{eq:lien23}
a_2 = - b_2 \ \ \ \mbox{and}\ \ \ a_3 = 4 b_2^2 -2 b_3.
\eea
Contrarily to the fermionic case, the cumulant $b_3$ and the virial coefficient $a_3$ of the bosons in the unitary
limit are not pure numbers, they rather are not yet explicitly determined functions of the three-body parameter.
After the pioneering works \cite{PaisUlhenbeck1959}, there exist for sure formal expressions of $b_3$ in the general 
quantum case involving Faddeev equations  \cite{GibsonReiner}, the $S$-matrix \cite{Ma}, Mayer diagrams \cite{Royer},
an Ursell operator \cite{Huang,Laloe1995}. The most operational form seems to be the one of \cite{GibsonReiner}
written in terms of a three-body scattering amplitude \cite{BedaqueRupak2003},
but its evaluation for the unitary Bose gas is, to our knowledge, purely numerical, and for three different values
of $R_t$ only \cite{BedaqueRupak2003}.
Here, we show on the contrary that $b_3$ can be obtained analytically, from the solution of the harmonically trapped
three-boson problem \cite{Pethick,Werner}.

In the first stage of the solution, one considers the system at thermal equilibrium in the harmonic potential
$U(\rr)=\frac{1}{2} m \omega^2 r^2$, $\omega$ being the free oscillation angular frequency, and one expresses
the third cumulant $B_3(\omega)$ of the trapped gas, defined by the expansion (\ref{eq:dev_piege}) to come,
in terms of the partition functions of $n$-body problems in the trap, $1\leq n\leq 3$.
In theoretical physics, a formal ``harmonic regularisator" was already introduced to obtain the
second \cite{Ouvry} and third \cite{Cabe,Law} virial coefficients of a gas of anyons.
The same technique was then used in the case of cold fermionic atoms  \cite{Drummond}, to take advantage
of the fact that the three-body problem is solvable in the unitary limit; note that it is then not a pure calculation
trick anymore since the trapping can be realised experimentally.

In the second stage, one takes the limit of a vanishing trap spring constant. As shown by the local density
approximation \cite{Drummond}, which is actually exact in that limit, the cumulants $b_n$ of the homogeneous gas
are then given by \cite{Ouvry,Drummond}
\be
\label{eq:lien}
\frac{b_n}{n^{3/2}}  = \lim_{\omega\to 0} B_3(\omega) \equiv B_3(0^+).
\ee
In \cite{Drummond}, this limit is evaluated purely numerically for the fermions. We shall insist here on showing
that one can actually go much farther analytically.

\section{Expression of $\Delta B_3$ in terms of canonical partition functions}

When the chemical potential $\mu\to -\infty$ at fixed temperature, which corresponds to a low density
limit, the grand potential of the trapped unitary Bose gas can be expanded as
\be
\Omega=- k_B T Z_1 \sum_{n\geq 1} B_n z^n
\label{eq:dev_piege}
\ee
where $Z_N$ is the canonical partition function with $N$ particles, $z=\exp(\beta \mu)$ is the fugacity,
and $B_n(\omega)$ is the $n^{\mathrm{th}}$ cumulant of the trapped gas. By construction, $B_1(\omega)\equiv 1$.
One also has, by definition, 
$\Omega = -k_B T \ln (1+\sum_{N\geq 1} Z_N z^N).$
By order-by-order identification in $z$, as for example in \cite{Drummond}, one finds
\be
B_2=\frac{Z_2}{Z_1} -\frac{1}{2} Z_1 \ \ \ \mbox{and}\ \ \ B_3=\frac{Z_3}{Z_1} -Z_2 + \frac{1}{3} Z_1^2.
\ee
In reality, one shall calculate the deviation from the ideal gas,
\be
\label{eq:defdel}
\Delta B_n  = B_n - B_n^{(0)}
\ee
where $B_n^{(0)}(\omega)$ is the $n^{\mathrm{th}}$ cumulant of the trapped ideal gas. Similarly one introduces
$\Delta Z_N = Z_N-Z_N^{(0)}$, noting that $\Delta Z_1=0$.
As there is furthermore separability of the center of mass in a harmonic trap, both for the ideal gas and for the unitary gas,
and as the spectrum of the center of mass is the same as the one of the one-body problem, one is left with partition
functions of the $n$-body relative motion, denoted by ``rel":
\bea
\label{eq:def_db2}
\Delta B_2 &=& \frac{\Delta Z_2}{Z_1} = \Delta Z_2^{\rm rel}\\
\label{eq:def_db3}
\Delta B_3 &=& \frac{\Delta Z_3}{Z_1}-\Delta Z_2 = \Delta Z_3^{\rm rel} - \Delta Z_2.
\eea
It remains to use the expressions for the relative motion spectrum, known in the unitary limit up to $n=3$  \cite{Pethick,Werner}.

\subsection{Case $N=2$}
The interactions modify the spectrum of the relative motion only in the sector of angular momentum $l=0$,
and in the unitary limit, their effect reduces to a downward shift by $\hbar \omega$ of the unperturbed spectrum,
$E_{l=0,n}^{(0)}=(2n+3/2)\hbar \omega$ \cite{Wilkens}, so one finds
\be
\Delta Z_2^{\rm rel} = (e^x-1) \sum_{n\geq 0} e^{-x(2n+3/2)}= \frac{e^{-x/2}}{1+e^{-x}}=\Delta B_2
\ee
where $x=\beta \hbar \omega$. As a consequence, using equation (\ref{eq:def_db2}) and $Z_1=
[e^{-x/2}/(1-e^{-x})]^3$:
\be
\label{eq:dz2}
\Delta Z_2 = \frac{e^{-2x}}{(1-e^{-x})^2(1-e^{-2x})}. 
\ee
Another consequence is that the second cumulant of the spatially homogeneous unitary Bose gas is obtained
from the expression $B_2^{(0)}(\omega)=\frac{1}{2} (\frac{1}{2\ch (x/2)})^3$ (deduced by a direct calculation)
by taking the $\omega\to 0$ limit and using (\ref{eq:lien}):
\be
\label{eq:b2unit}
\frac{b_2}{2^{3/2}} = \frac{1}{2} + \frac{1}{2^4}.
\ee

\subsection{Case $N=3$}
As shown in \cite{Werner}, the problem of three trapped bosons has a particular class of eigenstates, called 
laughlinian in what follows, 
with a wavefunction that vanishes when there is at least two particles in the same point.
These eigenstates thus have energies that are independent of the scattering length $a$ and they do not contribute to $\Delta Z_3$. It therefore remains to sum over the non-laughlinian eigenstates for the non-interacting case ($a=0$) 
and for the unitary limit ($1/a=0$).

In the case $a=0$, using as in \cite{Drummond} the Efimov ansatz for the three-body wavefunction
\footnote{\label{note:laugh} The non-laughlinian states of the ideal gas are indeed the limit for $a\to 0^-$ of the 
non-laughlinian states of the interacting case, for which a Faddeev type ansatz for the wavefunction is justified, 
$\psi(\rr_1,\rr_2,\rr_3)=
(1+P_{13}+P_{23}) \mathcal{F}(r,\rhob,\mathbf{C})$, where $P_{ij}$ transposes particles $i$ and $j$,
$\rr=\rr_2-\rr_1$, $\rhob=(2\rr_3-\rr_1-\rr_2)/\sqrt{3}$ and $\mathbf{C}=(\rr_1+\rr_2+\rr_3)/3$. This can be seen formally by integrating
Schr\"odinger's equation for a regularized-$\delta$ interaction, in terms of the Green's function of the non-interacting 
three-body Hamiltonian.}
one finds that the non-laughlinian eigenenergies of the relative motion are 
\be
\label{eq:ener0}
E_{\rm rel}^{(0)} = (u_{l,n}^{(0)}+1 + 2 q) \hbar \omega \ \ \ \mbox{of degeneracy}\ 2l+1,
\ee
where $l,n,q$ span the set of natural integers, with $l$ quantifying the angular momentum and 
$q$ the excitation of the hyperradial mode \cite{WernerSym}. The $u^{(0)}\geq 0$ are the roots of the function 
$s\mapsto 1/\Gamma(\frac{l-s}{2}+1)$~:
\be
u_{l,n}^{(0)} = l+2+2n.
\ee
In reality, one must keep in the spectrum (\ref{eq:ener0}) the physical roots only, such that the Efimov ansatz is not identically zero. The two known unphysical roots are $u^{(0)}=4$ for $l=0$, and $u^{(0)}=3$ for $l=1$. But this will not play any role 
in what follows.

In the case $1/a=0$,  using the Efimov ansatz as in \cite{Werner}, subjected to the Wigner-Bethe-Peierls contact conditions,
one gets the transcendental equation $\Lambda_l(s)=0$, with \cite{GasaneoBirse}
\begin{multline}
\label{eq:hypergeom}
\Lambda_l(s)=\cos\nu - (-\sin\nu )^l \frac{\Gamma(\frac{l+1+s}{2})\Gamma(\frac{l+1-s}{2})}{\pi^{1/2}\Gamma(l+\frac{3}{2})} \\
\times
{}_2 F_1\left(\frac{l+1+s}{2}, \frac{l+1-s}{2},l+\frac{3}{2}; \sin^2 \nu\right)
\end{multline}
where ${}_2 F_1$ is the Gauss hypergeometric function and the usual mass angle what introduced,
which is given for three bosons by
\be
\label{eq:nu}
\nu = \arcsin \frac{1}{2} = \frac{\pi}{6}.
\ee
In practice, we shall also use the representation derived in \cite{Tignone} for fermions and easily transposed to the
bosonic case, in terms of Legendre polynomials $P_l(X)$~:
\bea
\label{eq:rep_even}
\!\!\!\Lambda_l(s)\!\! &\!\stackrel{l\, {\rm even}}{=}\!& \!\!\cos \nu - \frac{2}{\sin \nu}\!\! \int_0^{\nu} \!\! d\theta
P_l\left(\frac{\sin\theta}{\sin \nu}\right) \frac{\cos (s\theta)}{\cos(s\pi/2)} \\
\!\!\!\Lambda_l(s)\!\! &\!\stackrel{l\, {\rm odd}}{=}\!&\!\! \cos \nu + \frac{2}{\sin \nu}\!\! \int_0^{\nu}\!\!  d\theta
P_l\left(\frac{\sin\theta}{\sin \nu}\right) \frac{\sin (s\theta)}{\sin(s\pi/2)},
\label{eq:rep_odd}
\eea
which allows us to explicitly write $\Lambda_l(s)$ with the sine function and rational fractions of $s$.

\subsection{Efimovian channel}
In the zero-angular-momentum sector, $l=0$, $\Lambda_l(s)$ has one and only one root $u_{0,0}\in i \mathbb{R}^+$ \cite{Werner}, 
usually noted as 
\be
\label{eq:s0ex}
u_{0,0} = s_0 = i |s_0|, \ \ |s_0|=1,006\, 237\, 825\, \ldots
\ee
This root gives rise to the efimovian channel, where the eigenenergies $\epsilon_q(\omega)$ of the relative motion
solve a transcendental equation 
\cite{Pethick,Werner} that one can rewrite as in \cite{Werner_gen2} to make explicit and univocal the dependence with
the quantum number $q\in\mathbb{N}$~:
\be
\label{eq:trans}
\im \ln \Gamma\Big(\frac{1+s_{0}-\epsilon_q/(\hbar\omega)}{2}\Big)
+\frac{|s_{0}|}{2} \ln \Big(\frac{2\hbar\omega}{E_t}\Big) + q \pi =0,
\ee
the function $\ln \Gamma(z)$ being taken with its standard determination (branch cut in  $\mathbb{R}^-$). 
In the limit $\omega\to 0$ for a fixed $q$, this reproduces the geometric sequence of Efimov trimers:
\be
\label{eq:trimlibre}
\epsilon_q(\omega) \to \epsilon_q(0^+) = -e^{-2\pi q/|s_0|} E_t,
\ee
which shows that $E_t =2\exp[\frac{2}{|s_0|}\im \ln \Gamma(1+s_0)]\hbar^2/(m R_t^2)$, $R_t$ being the three-body parameter 
according to the convention of \cite{Werner}.
In a strict zero-range limit, $q$ would span $\mathbb{Z}$ and the spectrum would be unbounded below, which would
prevent thermal equilibrium
of the system. As noted by Efimov \cite{Efimov}, however, 
in any given interaction model of finite range $b$, including experimental reality, the geometric form (\ref{eq:trimlibre})
of the spectrum only applies to the trimers of binding energy much smaller than $\hbar^2/(m b^2)$,
possible more deeply bound trimers being out of the unitary limit and non universal.
Here, the quantum number $q=0$ thus simply corresponds to the first state having (almost) reached the unitary limit.
For an interaction to allow for the realisation of the unitary Bose gas at thermal equilibrium,
$q=0$ must correspond to the {\sl true} ground trimer, and this is indeed the case for the model of \cite{vonStecher}
and for the narrow Feshbach resonance \cite{Petrov_bosons,Gogolin,Pricoupenko2010,Zwerger}:
in both situations, $E_t$ is indeed of the order of $e^{-2\pi/|s_0|} \hbar^2/(mb^2)
\ll \hbar^2/(mb^2)$ and the trimer spectrum can be considered as being {\sl entirely} efimovian, as it is assumed in the present work.

\subsection{Universal channels}
The real positive roots $(u_{0,n})_{n\geq 1}$ of $\Lambda_{0}(s)$, and the roots $(u_{l,n})_{n\geq 0}$ of $\Lambda_l(s)$
for $l>0$, which are all real \cite{Werner} and taken in what follows to be positive, 
give rise to universal, that is non-efimovian, states,
with eigenenergies of the relative motion that are independent of $R_t$:
\be
\label{eq:ener}
E_{\rm rel}=(u_{l,n}+1 + 2 q) \hbar \omega \ \ \ \mbox{of degeneracy}\ 2l+1,
\ee
where $(l,n)$ spans $\mathbb{N}^{2*}$, and $q$ spans $\mathbb{N}$, and the star indicates
that the vanishing element [here $(0,0)$] has to be excluded.
One can note the similitude with the non-interacting case (\ref{eq:ener0}), the quantum number $q$ having the
same physical origin \cite{Werner}.
As in the non-interacting case, one must eliminate from the spectrum (\ref{eq:ener}) the unphysical roots $u$,
that give a vanishing Efimov ansatz. These unphysical roots, however, are exactly the same ones in both cases
\footnote{For $a=0$, one requires that the wavefunction $\psi$ does not diverge as $1/r$ when $\rr=\rr_2-\rr_1\to \mathbf{0}$.
For $1/a=0$, one requires that $\psi$ has no $r^0$ term in its expansion in powers of $r$ (for fixed $\rr_1+\rr_2$ and $\rr_3$). 
When $\psi\equiv 0$, both constraints are satisfied, in which case $u^{(0)}$ and $u$ coincide.},
which allows us formally to include them in the partition functions 
$Z_3$ et $Z_3^{(0)}$, since their (unphysical) contributions exactly compensate in $\Delta Z_3$.
Collecting the contributions of the systems with and without interaction of common quantum numbers, we finally obtain:
\begin{multline}
\label{eq:dz3_rel_brut}
\Delta Z_3^{\rm rel} = \sum_{q\geq 0} \left[e^{-\beta \epsilon_q(\omega)}-e^{-x(u_{0,0}^{(0)}+1+2q)}\right] \\
\!\!\!\!+ \sum_{(l,n)\in \mathbb{N}^{2*}} \sum_{q\geq 0} (2l+1) \left[e^{-x(u_{l,n}+1+2q)}-e^{-x(u_{l,n}^{(0)}+1+2q)}\right].
\end{multline}

\subsection{Some useful transforms}
In order to treat one by one the problems that arise in taking the limit $\omega\to 0$, it is useful to split
(\ref{eq:dz3_rel_brut}) as the sum of a purely efimovian contribution $S(\omega)$ and of a purely universal contribution
$\sigma(\omega)$, up to additive remainders $R(\omega)$ and $\rho(\omega)$. In what follows we shall study
the efimovian series
\be
S(\omega) \equiv \sum_{q\geq 0} \left[e^{-\beta\epsilon_q(\omega)} -e^{-2qx}\right],
\ee
that reproduces the first sum in (\ref{eq:dz3_rel_brut}) up to the remainder
\be
\label{eq:Rom}
R(\omega) =\sum_{q\geq 0} \Big[e^{-2qx}-e^{-x (u_{0,0}^{(0)}+1+2q)}\Big] = \frac{1-e^{-3x}}{1-e^{-2x}}.
\ee
We shall see that it is a doubly clever idea to introduce the universal series
\be
\label{eq:sigma}
\sigma(\omega)=\!\!\!\sum_{(l,n)\in \mathbb{N}^{2*}} \sum_{q\geq 0} (2l+1) \Big[e^{-x (u_{l,n}+1+2q)} - e^{-x(v_{l,n}+1+2q)}\Big],
\ee
with
\be
v_{l,n}=l+1+2n.
\ee
First, this provides a numerical advantage \cite{Drummond}, since $v_{l,n}$ is the large-$l$-or-$n$ equivalent of $u_{l,n}$ 
introduced in equation (17) of reference \cite{Werner}\footnote{That equation (17) for $l<2$ contains an error, 
induced by a wrong labeling of the unphysical roots.}, so that the series $\sigma$ rapidly converges.
Second, as it is apparent in the form (\ref{eq:hypergeom}), the $(v_{l,n})_{n\geq 0}$ are the positive poles of the function $\Lambda_l(s)$; the writing (\ref{eq:sigma}) is thus reminiscent
of the residue theorem, and we shall soon take advantage of this fact. As $v_{l,n}=u_{l,n}^{(0)}-1$, $\sigma$
reproduces the second sum of (\ref{eq:dz3_rel_brut}) up to the remainder
\be
\rho(\omega)=\!\!\!\sum_{(l,n)\in\mathbb{N}^{2*}} \sum_{q\geq 0} (2l+1) 
\Big[e^{-x(v_{l,n}+1+2q)}-e^{-x({u}_{l,n}^{(0)}+1+2q)}\Big].
\ee
With the generating function method, one gets
$\sum_{l\geq 0} (2l+1) e^{-lx}=(1-2\frac{d}{dx})\frac{1}{1-e^{-x}}= (1+e^{-x})/(1-e^{-x})^2$.
This, together with the identities (\ref{eq:dz2}) and (\ref{eq:Rom}), leads to $\rho(\omega)=\Delta Z_2+1-R(\omega)$,
and (\ref{eq:def_db3}) reduces to the simple writing
\be
\label{eq:db3fin}
\Delta B_3(\omega) = S(\omega)+\sigma(\omega)+1.
\ee
This equation  (\ref{eq:db3fin}) is the bosonic equivalent of the fermionic expressions (56,58) of reference
\cite{Drummond}, from which it differs mainly through the contribution $S(\omega)$ of the efimovian channel.

\section{Analytical transforms and $\omega\to 0$ limit}
The most relevant physical quantity being the third cumulant of the spatially homogeneous gas, we must now,
according to (\ref{eq:lien}), take the limit of a vanishing trap spring constant. An explicit calculation for
the trapped ideal gas gives
\be
\label{eq:B30}
B_3^{(0)}(\omega) =\frac{1}{3} \left(\frac{e^{-x}(1-e^{-x})}{1-e^{-3x}}\right)^3\underset{\omega\to 0}{\to} \frac{1}{3^4}.
\ee
In the unitary case, the method of reference \cite{Werner_gen2} allows one to determine $S(0^+)$ exactly;
furthermore, as we shall see, the sum over $n$ in equation (\ref{eq:sigma}) and the $x\to 0$ limit
can be performed analytically, and the resulting integrals in $\sigma(0^+)$, that are very simple
to evaluate numerically, can also be usefully evaluated by taking the mass angle $\nu$ as a small parameter.

\subsection{Working on the efimovian contribution  $S(\omega)$}
For an arbitrary positive number $A$, with $A\gg 1$, we split as in \cite{Werner_gen2} the series (\ref{eq:sigma}) 
into three pieces, $S=S_1+S_2+S_3$.

In the first (quasi-bound) piece, that contains the states such that $\epsilon_q(\omega) < -A \hbar \omega$, 
the three bosons occupy a spatial zone much smaller than the ground-state harmonic oscillator length $[\hbar/(m\omega)]^{1/2}$,
so that the spectrum is close to the free-space spectrum (\ref{eq:trimlibre}):
\be
\epsilon_q(\omega)=\epsilon_q(0^+)\Big[1-\frac{1+|s_0|^2}{6} \Big(\frac{\hbar\omega}{\epsilon_q(0^+)}\Big)^2+\ldots\Big]
\ee
As $\epsilon_q(0^+)$ is a geometric sequence, approximating $\epsilon_q(\omega)$ with $\epsilon_q(0^+)$ in $S_1$
induces an error of the order of the error on the last term,  that vanishes as $x/A$.
From the same geometric property, the maximal index $q_1$ in that first piece
diverges only logarithmically, as $\frac{|s_0|}{2\pi} \ln[E_t/(2\hbar \omega)]$, which allows one to replace each term 
$e^{-2qx}$ by 1, the resulting error on $S_1$ vanishing as $x q_1^2$. One thus has
\be
S_1 = \sum_{q=0}^{q_1} \Big[e^{-\beta \epsilon_q(0^+)}-1\Big] + o(1).
\ee

The second piece contains the intermediate states, such that $|\epsilon_q(\omega)| < A\hbar\omega$. 
As the level spacing between the $\epsilon_q(\omega)$
is of order $\hbar\omega$ at least\footnote{The derivative with respect to $\epsilon_q/(\hbar \omega)$ 
of the left-hand side of equation (\ref{eq:trans}) is indeed uniformly bounded, according to relation 8.362(1) 
of reference \cite{Gradstein}.}, 
it contains a finite number $O(A)$ of terms, and each terms vanishes when $\omega$ vanishes, so that
$S_2=o(1)$.

The third piece contains the states $A<\epsilon_q(\omega)$, that reconstruct the free space efimovian scattering states
when $\omega\to 0$.  On can use for these states the large-$q$ expansion 
\be
\frac{\epsilon_q(\omega)}{\hbar\omega}= 2q+\Delta\big(\epsilon_q(\omega)\big)+O(1/q)
\ee
where the dimensionless function of the energy $\Delta(\epsilon)$ is given by equation (C6) of \cite{Werner_gen2}.
In each term of $S_3$, one uses the approximation $e^{-2qx}\simeq e^{-\beta\epsilon_q}[1+x\Delta(\epsilon_q)]$,
and one replaces the sum over $q$ by an integral over energy; since the level spacings are almost constant, 
$\epsilon_{q+1}-\epsilon_q=2\hbar \omega[1+O(\hbar\omega/\epsilon_q)]$ \cite{WernerThese}, one gets
\be
S_3= -\frac{1}{2} \int_{A\hbar \omega}^{+\infty}   d\epsilon  \, e^{-\beta\epsilon}  \beta \Delta(\epsilon) +o(1).
\ee
Collecting the three pieces and writing up $\Delta(\epsilon)$ explicitly as in \cite{Werner_gen2},
one finally obtains for $\omega\to 0$:
\begin{multline}
\label{eq:defS0}
S(0^+)=
\Big\{\sum_{q\geq 0} \left[e^{-\beta \epsilon_q(0^+)}-1\right] \Big\}
+\frac{|s_{0}|}{\pi}\Big\{\frac{1}{2}\ln (e^\gamma \beta E_t)  \\
-\sum_{p\geq 1} \! e^{-p\pi |s_{0}|} \re\!\Big[\Gamma(-ip|s_{0}|) (\beta E_t)^{i p |s_{0}|}\Big]\!\Big\},
\end{multline}
where $\gamma=0.577\, 215\ldots$ is Euler's constant, and the free space trimer energy 
$\epsilon_q(0^+)$ is given by (\ref{eq:trimlibre}). One can note that the bound state contribution
$\sum_{q\geq 0} e^{-\beta \epsilon_q(0^+)}$ is divergent, and has thus to be collected with the contribution of the continuum 
to obtain the counter-term $-1$ ensuring the convergence of the sum in (\ref{eq:defS0});
in the case of the second virial coefficient of a plasma,
the (two-body) bound states have an hydrogenoid spectrum, which requires the more elaborated counter-term 
$-(1+\beta \epsilon_q)$ to get a converging sum \cite{Larkin}\footnote{Changing $R_t$ into $\tilde{R}_t=e^{-\pi/|s_0|}R_t$,
according to (\ref{eq:trans}), has the only effect on the efimovian spectrum of adding a  ``$q=-1$" state, 
so that $\tilde{S}(0^+)=
\exp(\beta E_t e^{2\pi/|s_0|})+S(0^+)$. This functional property determines $S(0^+)$ 
[and reproduces the first two contributions in (\ref{eq:defS0})] up to an unknown additive function of
$\ln(\beta E_t)$ with a period $2\pi/|s_0|$.}.

\subsection{Working on the universal contribution $\sigma(\omega)$}
Let us extract from the definition (\ref{eq:sigma}) of $\sigma(\omega)$ the contribution of the angular momentum $l$
and let us sum over $q$, to obtain
\be
\sigma=\sum_{l\geq 0} \sigma_l,\ \ \mbox{with}\ \ 
\sigma_l= \frac{l+\frac{1}{2}}{\sh x} \sum_{n\geq \delta_{l,0}} \Big(e^{-x u_{l,n}} - e^{-x v_{l,n}}\Big).
\ee
The key point is now that the function $\Lambda_l(s)$ has a simple root\footnote{\label{note:racs}
To prove by contradiction the absence of multiple roots,  let us recall that
the hyperangular part of the Efimov ansatz is $\Phi(\Omegab)=(1+P_{13}+P_{23}) \frac{\varphi(\alpha)}{\sin 2\alpha}
Y_l^{m_l}(\rhob/\rho)$, where we introduced, in additions to the notations of footnote [52], 
the variable $\alpha =\atan(r/\rho)$, the spherical harmonics
$Y_l^{m_l}$, the set $\Omegab$ of the five hyperangles and a function $\varphi(\alpha)$ that vanishes in $\pi/2$. 
Schr\"odinger's equation then imposes 
$(4-s^2-\Delta_{\Omegabs}) \Phi=0$, where $\Delta_{\Omegabs}$ is the Laplacian on the hypersphere, so that
$-\varphi''+\frac{l(l+1)}{\cos^2\alpha}\varphi=s^2 \varphi$. The Wigner-Bethe-Peierls contact conditions for 
$1/a=0$ further impose the boundary condition
$\varphi'(0)+\frac{4(-1)^l}{\cos\nu} \varphi(\frac{\pi}{2} -\nu)=0$, which is essentially the transcendental 
equation $\Lambda_l(s)=0$ [see footnote [58]]. 
If $s>0$ is a double root, $\psi(\alpha)=\partial_{s^2} \varphi(\alpha)$ obeys the same boundary 
conditions as $\varphi(\alpha)$, with $-\psi''+\frac{l(l+1)}{\cos^2\alpha}\psi=s^2 \psi+\varphi$. Then
$\Psi(\Omegab)\equiv (1+P_{13}+P_{23}) \frac{\psi(\alpha)}{\sin 2\alpha} Y_l^{m_l}(\rhob/\rho)$ obeys
the $1/a=0$ Wigner-Bethe-Peierls contact conditions, and the inhomogeneous equation
$(4-s^2-\Delta_{\Omegabs}) \Psi=\Phi$; the scalar product of this equation which $\Phi$ leads to the
contradiction $0=\int d^5\Omega |\Phi(\Omegab)|^2$, 
since the Laplacian is hermitian for fixed contact conditions.
}
in $u_{l,n}$ and a simple pole\footnote{\label{note:polef} For particular values of $\nu$ different from $\pi/6$, 
{\sl e.g.\ } $\nu=\pi/5$,
the function ${}_2 F_1$ in
(\ref{eq:hypergeom}) vanishes in $s=v_{l,n}$, in which case $v_{l,n}$ is not a pole of $\Lambda_l(s)$.
The true form of the transcendental equation for $s$, however, as given in footnote [57] 
and whose the $u_{l,n}$ must be the roots,
is $\Lambda_l(s)/[\Gamma(\frac{l+1+s}{2})\Gamma(\frac{l+1-s}{2})]=0$; there is then a fully acceptable root 
$u_{l,n}=v_{l,n}$, 
which is however not a root of $\Lambda_l(s)$ and was thus missed by the form  (\ref{eq:hypergeom}).
As these ``ghost" roots and poles of $\Lambda_l(s)$ actually coincide, they do not contribute to $\sigma_l(\omega)$.}
in $v_{l,n}$, so that its logarithmic derivative has a pole in both points, with a residue respectively
equal to $+1$ et $-1$.
By inverse application of the residue formula, one thus finds for $l>0$ that
\be
\sigma_l(\omega) \stackrel{l>0}{=} \frac{l+\frac{1}{2}}{\sh x} \int_C \frac{dz}{2i\pi} \frac{\Lambda_l'(z)}{\Lambda_l(z)} e^{-xz}
\ee
where the integral is taken over the contour $C$ that comes from  $z=+\infty+i\eta$ ($\eta>0$), 
moves parallelly to and above the real axis, crosses the real axis close to the origin and then tends to
$z=+\infty-i\eta$ moving parallelly to and below the real axis. This contour indeed encloses all the positive roots
$(u_{l,n})_{n\geq 0}$ and all the positive poles $(v_{l,n})_{n\geq 0}$ of the function $\Lambda_l(z)$.
As $\Lambda_l(z)$ ($l>0$) has no other roots or poles in the half-plane $\re{z}\geq 0$,
one can unfold $C$ around the origin and map it to the purely imaginary axis $z=iS$:
\be
\sigma_l(\omega) \stackrel{l>0}{=} \frac{l+\frac{1}{2}}{\pi\sh x } \int_{0}^{+\infty} dS  \frac{\Lambda_l'(iS)}{\Lambda_l(iS)} i\sin (xS)
\ee
where the fact that $\Lambda_l'(iS)/\Lambda_l(iS)$ is an odd function allows one to omit the $\cos(xS)$ and to restrict integration to  $S>0$.
It is then elementary to take the $x\to 0$ limit, and a simple integration by parts leads to the nice result:
\be
\label{eq:sigl0}
\sigma_l(0^+) \stackrel{l>0}{=} -\frac{2l+1}{2\pi}  \int_{0}^{+\infty} dS \ln\Big(\frac{\Lambda_l(iS)}{\cos \nu}\Big).
\ee
According to (\ref{eq:rep_even},\ref{eq:rep_odd}), the function $\Lambda_l(iS)/\cos\nu$ exponentially tends to 1
at infinity, so that the integral in (\ref{eq:sigl0}) rapidly converges.

The case $l=0$ requires some twist of the previous reasoning. First, the pole  $v_{0,0}$ of the function 
$\Lambda_0(s)$ does not contribute to $\sigma(\omega)$, since $(l,n)=(0,0)$ is in the efimovian channel.
Second, the existence of the efimovian roots $\pm i|s_0|$ of $\Lambda_0(s)$ prevents one from
folding back the integration contour $C$ on the purely imaginary axis. 
Both points are solved by considering the function $\frac{s^2-v_{0,0}^2}{s^2-s_0^2}\Lambda_0(s)$ rather than
the function $\Lambda_0(s)$ itself: The rational prefactor suppresses the poles $\pm v_{0,0}$ and
the roots $\pm s_0$ without destroying the parity invariance that we have used.
In the limit $\omega\to 0$, this leads to\footnote{A related integral appears in equation (87) of \cite{Tignone},
which establishes an unexpected link between $b_3$ and the value of $R_t/R_*$ on a narrow Feshbach
resonance (with a Feshbach length $R_*$).}
\be
\label{eq:sig00}
\sigma_0(0^+) = -\frac{1}{2\pi} \int_{0}^{+\infty} dS \ln\Big(\frac{S^2+1}{S^2-|s_0|^2}\,\frac{\Lambda_0(iS)}{\cos \nu}\Big).
\ee
From the writing  (\ref{eq:rep_even},\ref{eq:rep_odd}) of the functions $\Lambda_l(s=iS)$, and using
the numerical tools of formal integration software, one obtains
in a few minutes, 
the value of the constant term in $\Delta B_3(0^+)$:
\be
\label{eq:ctex}
1+\sigma(0^+)= 1-0,364\, 037\, \ldots= 0,635\, 962\, \ldots
\ee
One observes that $(\sigma_l(0^+))_{l\geq 1}$ is an alternating sequence with a rapidly decreasing modulus,
so that the error due to a truncation in $l$ is bounded by the first neglected term.
Analytically, one can obtain the elegant asymptotic form\footnote{In (\ref{eq:hypergeom}), 
one replaces ${}_2 F_1$ by the usual series expansion, involving a sum over $k\in\mathbb{N}$,
see relation 9.100 in reference \cite{Gradstein}. 
One takes the limit $l\to +\infty$ for fixed $y\equiv k/l$ and $\tau\equiv S/l^{1/2}$,
one approximates each term by its Stirling equivalent, and one replaces
the sum over $k$ by an integral over $y$.
The resulting integrand contains a factor 
$e^{l u(y)}$, where $u(y)=2(y+\frac{1}{2})\ln(y+\frac{1}{2})-(y+1)\ln(y+1)-y\ln y
+2y\ln (\sin\nu)$, which allows one to use Laplace method and leads to
$\delta_l(S)\sim \frac{(-1)^{l+1} 2^{1/2}}{l\cos\nu \cos\frac{\nu}{2}} (\tan\frac{\nu}{2})^l \exp(-\frac{1}{2}\tau^2 \cos\nu)$
[a quantity defined in (\ref{eq:delta})], whose insertion in (\ref{eq:sigl0}) gives (\ref{eq:grandl}).}
\be
\label{eq:grandl}
\sigma_l(0^+) \underset{l\to \infty}{\sim}  \Big(\frac{l}{\pi}\Big)^{1/2} \frac{(-\tan \frac{\nu}{2})^l}{\cos \frac{\nu}{2} (\cos \nu)^{3/2}},
\ee
where $\nu$ is the mass angle (\ref{eq:nu}).

\subsection{An entirely analytical evaluation}
The fact that $\sigma_l(0^+)$ is rapidly decreasing with the angular momentum, even in the absence 
of physical interpretation, can be understood from the fact that, for $l>0$, the deviations 
of $\Lambda_l(iS)/\cos\nu$ from unity vanish as
$-(-\nu)^l$, this is obvious on the writing  (\ref{eq:hypergeom}):
\be
\label{eq:delta}
\delta_l(S) \equiv \frac{\Lambda_l(iS)}{\cos\nu}-1 \underset{\nu\to 0}{=} O (\nu^l).
\ee
For $\nu=\pi/6$, one finds that, already for $l=1$, the maximum of $|\delta_l(S)|$, reached in $S=0$,
has the small value $\simeq 0,273$. This gives the idea of treating each $\delta_l$ (for $l>0$) as an infinitesimal
quantity of order $l$. The series expansion
of $\ln[1+\delta_l(S)]$ in powers of $\delta_l$ in (\ref{eq:sigl0}) is convergent and generates a convergent 
expansion of $\sigma_l(0^+)$~:
\be
\sigma_l(0^+) \stackrel{l>0}{=}\sum_{n\geq 1} \sigma_l^{(n)}, \ \mbox{with}\ 
\sigma_l^{(n)}=(2l+1) \frac{(-1)^n}{n} \int_{\mathbb{R}} \frac{dS}{4\pi} [\delta_l(S)]^n.
\ee
The resulting integral can in principle be calculated analytically by the residue formula, for $0<\nu<\pi/2$,
which leads to series that can be expressed in terms of the Bose functions 
$g_\alpha(z)=\sum_{k\geq 1} z^k/k^\alpha$, also called polylogarithms, but this rapidly becomes
tedious at large $l$ or $n$.
We thus restrict ourselves to the order 3 included. For $n\leq 2$, it is actually simpler to directly calculate 
the sum over all $l\geq 1$ of $\sigma_l^{(n)}$, that we note as $\sigma_{1:\infty}^{(n)}$. We finally keep 
as the desired approximation:
\be
\label{eq:sigapp}
\sigma(0^+) \approx \sigma_0(0^+) + \sigma_{1:\infty}^{(1)}+\sigma_{1:\infty}^{(2)}+\sigma_1^{(3)}.
\ee
The second and third terms of the approximation (\ref{eq:sigapp}) can be expressed simply for an arbitrary $\nu$ as
\footnote{This results, among other things, from equation (35) of \cite{Tignone} adapted to the bosonic case.  
In (\ref{eq:sig1}), this equation led to an expression of $\sigma_l^{(1)}$ in terms of the associated Legendre 
function $Q_l(X)$, that one writes as in relation
8.821(3) of \cite{Gradstein} to be able to sum over $l$. In (\ref{eq:sig2}), this equation was combined to the
Parseval-Plancherel identity (to integrate over $S$) and to the fact that the polynomials 
$(l+\frac{1}{2})^{1/2}P_l(X)$ form an orthonormal basis in the functional space
$L_2([-1,1])$ (to sum over $l$).}
\bea
\label{eq:sig1}
\!\!\!\!\!\sigma_{1:\infty}^{(1)} &\!\!\!=\!\!\!& \frac{1}{\pi\cos\nu\, (1+\sin \nu)}- \frac{\argth(\sin\nu)}{\pi\cos\nu\sin\nu}  \\
\!\!\!\!\!\sigma_{1:\infty}^{(2)} &\!\!\!=\!\!\!& \frac{2\nu}{\pi^2 \sin\nu \cos^3\nu}
- \frac{4[\frac{7}{8} \zeta(3) - \re C_3-\nu \im C_2]}{(\pi \sin \nu\cos \nu)^2}
\label{eq:sig2}
\eea
with $\zeta$ the Riemann function and $C_\alpha=g_\alpha(e^{2i\nu})-\frac{1}{2^\alpha} g_\alpha(e^{4i\nu})$. 
For $\nu=\pi/6$, one has simply  $\re C_3=\frac{7}{18}\zeta(3)$ and 
$\im C_2=\frac{\sqrt{3}}{72} [\psi'(\frac{1}{6})-\psi'(\frac{5}{6})]$,
where $\psi$ is the digamma function and $\psi'$ its first order derivative. To be concise,
we give the value of the last term of (\ref{eq:sigapp}) for $\nu=\pi/6$ only:
\begin{multline}
\sigma_1^{(3)}=
\frac{64}{\pi ^3 \sqrt{3}} \left(\frac{17 D_3}{432}-\frac{14 \zeta (3)}{3}-\frac{403 \zeta (5)}{27}+2\right) \\
+\frac{16}{9 \pi ^2} \left(\frac{17 D_3}{54}+5 D_1-\frac{322 \zeta (3)}{3}-36\right) \\
+\frac{32}{3\pi \sqrt{3}}  \left(\frac{5 D_1}{9}-\frac{112 \zeta (3)}{27}+\frac{8}{3}-2 \ln 3\right),
\end{multline}
with $D_k=\psi^{(k)}(\frac{1}{3})-\psi^{(k)}(\frac{2}{3})$, $\psi^{(k)}$ being the 
$k^{\mathrm{th}}$ derivative of the digamma function, see relation 8.363(8) of reference \cite{Gradstein}.

It remains to analytically evaluate the first term of (\ref{eq:sigapp}),
that is the universal contribution at zero angular momentum $\sigma_0(0^+)$ given by
(\ref{eq:sig00}). Since a series expansion of the logarithm around $1$ is not suited to this case,
we directly expand to second order in powers of the mass angle\footnote{this is justified even under the integral, since the integral over $S$ in (\ref{eq:sig00}) converges over a distance of order unity.}: according to (\ref{eq:rep_even}),
\be
\frac{\Lambda_0(iS)}{\cos\nu}   
=1-\frac{2}{\ch(S\pi/2)} - \frac{4}{3}\nu^2 \frac{1+S^2/4}{\ch(S\pi/2)} + O(\nu^4).
\ee
This first allows one to evaluate the efimovian root
\be
|s_0| = \theta + \frac{8\nu^2}{3\pi\sqrt{3}} (1+\theta^2/4)+O(\nu^4)
\ee
where $\theta=\frac{2}{\pi} \argch 2=0,838\, 401\ldots$, 
in a way that reproduces (for $\nu=\pi/6$) its exact value (\ref{eq:s0ex}) within one part per thousand.
Then, after a few applications of the residue formula, one obtains the desired approximation up to order 3 included:
\be
\sigma_0(0^+) \simeq -\frac{1+\theta^2}{8}-\frac{2\nu^2}{9\pi\sqrt{3}}\, \theta (1+\theta^2/4).
\ee
For $\nu=\pi/6$, our analytical approximation (\ref{eq:sigapp})
then leads to 
\be
1+\sigma(0^+) \simeq  1-0, 364 \, 613\, \ldots = 0, 635\, 386\, \ldots
\ee
which reproduces the exact value (\ref{eq:ctex}) within one part per thousand.
Such a precision is sufficient in practice, considering the present uncertainty on the measurements
of the equation of state of the ultracold atomic gases
\cite{Nascimbene,Navon,Zwierlein} and the fact that $b_3$ is the coefficient of a term that has to be small
in a weak density expansion.

\section{Conclusion}
We have shown that one can entirely analytically determine the third cumulant $b_3$
of the spatially homogeneous unitary Bose gas of three-body parameter $R_t$ and thermal de Broglie wavelength
$\lambda$: The result\footnote{Indeed $C=1+\sigma(0^+)+B_3^{(0)}(0^+)$, see (\ref{eq:defdel}), 
(\ref{eq:db3fin}), (\ref{eq:B30}) and (\ref{eq:ctex}).}
\be
\label{eq:resultat}
b_3 = 3\sqrt{3} [S(0^+) + C] \ \ \mbox{with}\ \ C=0, 648 \ldots
\ee
is the sum of a function $S(0^+)$ of $\lambda/R_t$ exactly given by equation (\ref{eq:defS0}),
and of a constant $C$; we have found an original integral representation of $C$ that makes its numerical evaluation
straightforward, and that allows one to perform a perturbative expansion of $C$, in principle to arbitrary order but
restricted here to third order included,
taking the mass angle $\nu=\pi/6$ as a small parameter.
This gives access to the third virial coefficient $a_3$ of the unitary Bose gas,
combining relations (\ref{eq:lien23}) and (\ref{eq:b2unit}).

\begin{figure}[t]
\includegraphics[width=8cm,clip=]{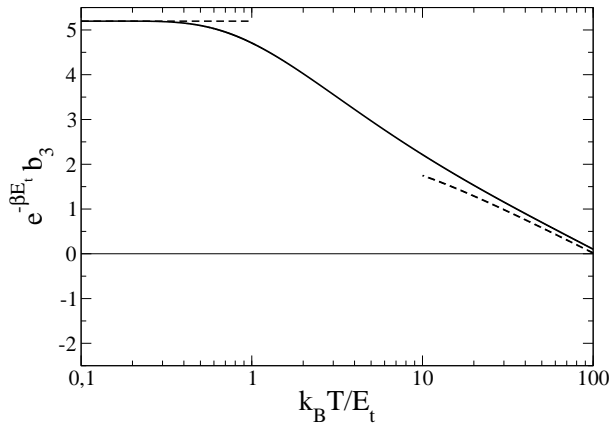}
\caption{Third cumulant $b_3$ of the spatially homogeneous unitary Bose gas: $b_3$ is multiplied by 
$e^{-\beta E_t}$ and plotted as a function of temperature $T$ as a solid line.
The dashed lined to the left and to the right respectively correspond to the approximations
(\ref{eq:equiv_exp}) and (\ref{eq:equiv_log}), that have been multiplied by $e^{-\beta E_t}$.
$(-E_t)$ is the energy of the ground state (efimovian) trimer, and $\beta=1/(k_B T)$.}
\end{figure}

As shown by the figure, our result has the physical interest of describing the crossover between two limiting regimes,
the low-temperature regime 
$k_B T \ll E_t$, where $b_3$ is dominated by the contribution of the ground state trimer of energy $-E_t$:
\be
\label{eq:equiv_exp}
b_3 \simeq 3\sqrt{3} \, e^{\beta E_t},
\ee
with $E_t\propto \hbar^2/(m R_t^2)$,
and the regime $k_BT \gg E_t$, where the trimers are almost fully dissociated:
\be
\label{eq:equiv_log}
b_3 \simeq 3\sqrt{3} \frac{|s_0|}{2\pi} \ln(e^{\gamma+2\pi C/|s_0|}\beta E_t).
\ee
The exponential approximation (\ref{eq:equiv_exp}) agrees with the expression (193) of reference \cite{PaisUlhenbeck1959},
that was very simply deduced from the chemical equilibrium condition of the gas.
The logarithmic approximation (\ref{eq:equiv_log}) can also be recovered, within a constant factor inside
the logarithm, by a calculation that totally differs from ours, the extraction of the loss rate constant  $L_3$
from the free-space inelastic scattering problem of three bosons \cite{SalomonPetrov},
further combined with equation (25) of \cite{Werner_gen2} that relates (through general arguments) 
${\partial b_3}/{\partial(\ln R_t)}$ to $L_3$ in the weak inelasticity limit of that scattering problem.

In practice, in the temperature range where the logarithmic approximation (\ref{eq:equiv_log}) well reproduces
our values of $b_3$, it may be difficult to ensure that the unitary limit is reached, {\sl i.e.\ }
that the finite (real or effective) range of the interaction is indeed negligible.
In particular, it is not guaranteed that the change of sign of $b_3$ at high temperature, 
as predicted by the zero-range efimovian theory used in this paper\footnote{Using the full expression 
(\ref{eq:resultat}), one finds that $b_3=0$ for $k_B T/E_t\simeq 112,56$.}, may really be observed
for a more realistic model such as the ones of
references \cite{vonStecher,Petrov_bosons,Gogolin,Pricoupenko2010},
or in ultracold-atom experiments.
Answering this question requires the study of a specific model for the interaction
and must be kept for future investigation.

\section*{Acknowledgments}
This work was performed in the frame of the ERC project FERLODIM. We thank our colleagues
of the ultracold fermion group at LKB, in particular Nir Navon, and we thank Xavier Leyronas at LPS,
for useful discussions about the unitary Bose gas and the virial coefficients.


\begin{thebibliography}{0}
\expandafter\ifx\csname natexlab\endcsname\relax\def\natexlab#1{#1}\fi
\expandafter\ifx\csname bibnamefont\endcsname\relax
  \def\bibnamefont#1{#1}\fi
\expandafter\ifx\csname bibfnamefont\endcsname\relax
  \def\bibfnamefont#1{#1}\fi
\expandafter\ifx\csname citenamefont\endcsname\relax
  \def\citenamefont#1{#1}\fi
\expandafter\ifx\csname url\endcsname\relax
  \def\url#1{\texttt{#1}}\fi
\expandafter\ifx\csname urlprefix\endcsname\relax\def\urlprefix{URL }\fi
\providecommand{\bibinfo}[2]{#2}
\providecommand{\eprint}[2][]{\url{#2}}

\end{thebibliography}


\begin{thebibliography}{99}


\bibitem{Heinzen}
J.M. Vogels {\sl et al.}, 
Phys.  Rev. A {\bf 56}, R1067 (1997).

\bibitem{Julienne}
C. Chin, R. Grimm, P. Julienne, E. Tiesinga, 
Rev. Mod. Phys. {\bf 82}, 1225 (2010).

\bibitem{prems}
K.M. O'Hara {\sl et al.}, 
Science {\bf 298}, 2179 (2002);
T. Bourdel {\sl et al.}, 
Phys. Rev. Lett. {\bf 91}, 020402 (2003);

\bibitem{BookZwerger}
{\sl The BCS-BEC Crossover and the Unitary Fermi Gas},
LNIP 836,
edited by W. Zwerger (Springer, Berlin, 2012).

\bibitem{Nascimbene}
S. Nascimb\`ene {\sl et al.}, 
Nature {\bf 463}, 1057 (2010).

\bibitem{Navon}
N. Navon, S. Nascimb{\`e}ne, F. Chevy, C. Salomon,
Science {\bf 328}, 729 (2010).

\bibitem{Zwierlein}
Mark J.H. Ku, A.T. Sommer, L.W. Cheuk, M.W. Zwierlein,
Science {\bf 335}, 563 (2012).

\bibitem{Gezerlis}
M. M. Forbes, S. Gandolfi, A. Gezerlis, 
Phys. Rev.  Lett. {\bf 106}, 235303 (2011).

\bibitem{Svistunov}
E. Burovski, N. Prokof'ev, B. Svistunov, M. Troyer,
Phys. Rev. Lett. {\bf 96}, 160402 (2006).

\bibitem{Goulko}
O. Goulko, M. Wingate, Phys. Rev. A {\bf 82}, 053621 (2010).

\bibitem{BDMC}
K. Van Houcke {\sl et al.}, 
Nature Phys. {\bf 8}, 366 (2012).

\bibitem{Pethick}
S. Jonsell,  H. Heiselberg, C.J. Pethick, Phys. Rev. Lett. {\bf 89}, 250401 (2002).

\bibitem{Drummond}
Xia-Ji Liu, Hui Hu, P. D. Drummond, Phys. Rev. Lett. {\bf 102}, 160401 (2009); Phys. Rev. A {\bf 82}, 023619 (2010).

\bibitem{Werner}
F. Werner, Y. Castin, Phys. Rev. Lett. {\bf 97}, 150401 (2006).

\bibitem{Leyronas}
X. Leyronas, 
Phys. Rev. A {\bf 84}, 053633 (2011).

\bibitem{Blume}
D. Rakshit, K. M. Daily, D. Blume,
Phys. Rev. A {\bf 85}, 033634 (2012).

\bibitem{Efimov}
V. Efimov,
Sov. J. Nucl. Phys. {\bf 12}, 589 (1971).

\bibitem{Revue}
F. Ferlaino, A. Zenesini, M. Berninger, B. Huang, H.-C. Nägerl, R. Grim,
Few-Body Syst. {\bf 51}, 113 (2011).

\bibitem{SalomonPetrov}
B. S. Rem {\sl et al.}, 
\url{http://hal.archives-ouvertes.fr/hal-00768038}

\bibitem{Esry}
J. P. D'Incao, H. Suno, B. D. Esry, 
Phys. Rev. Lett. {\bf 93}, 123201 (2004).

\bibitem{Werner_gen2}
F. Werner, Y. Castin,  Phys. Rev. A {\bf 86}, 053633 (2012).

\bibitem{Zoller}
A. J. Daley {\sl et al.}, 
Phys. Rev. Lett. {\bf 102}, 040402 (2009).

\bibitem{Petrov}
J. Levinsen, T.G. Tiecke, J.T.M. Walraven,  D.S. Petrov,
Phys. Rev. Lett. {\bf 103}, 153202 (2009).

\bibitem{Wang}
Yujun Wang, J. P. D'Incao, B.D. Esry, Phys. Rev. A {\bf 83}, 042710 (2011).

\bibitem{Pandha}
S. Cowell {\sl et al.}, 
Phys. Rev. Lett. {\bf 88}, 210403 (2002).

\bibitem{Stoof}
J. M. Diederix, T. C. F. van Heijst, H. T. C. Stoof,
Phys. Rev. A {\bf 84}, 033618 (2011).

\bibitem{Ho2012}
Weiran Li, Tin-Lun Ho,
Phys. Rev. Lett. {\bf 108}, 195301 (2012).

\bibitem{vonStecher}
J. von Stecher, J. Phys. B {\bf 43}, 101002 (2010).

\bibitem{Beth}
E. Beth, G.E. Uhlenbeck, Physica III 8, 729 (1936); Physica IV 10, 915 (1937).

\bibitem{Huang}
K. Huang, in {\sl Statistical Mechanics}, p. 427 (Wiley, New York, 1963).

\bibitem{PaisUlhenbeck1959}
A. Pais, G.E. Uhlenbeck, Phys. Rev. {\bf 116}, 250 (1959).

\bibitem{GibsonReiner}
W.G. Gibson, Phys. Letters {\bf 21}, 619 (1966);
A.S. Reiner, Phys. Rev. {\bf 151}, 170 (1966). 

\bibitem{Ma}
R. Dashen, Shang-keng Ma, H.J. Bernstein,
Phys. Rev. {\bf 187}, 345 (1969);
S. Servadio, Il Nuovo Cimento {\bf 102}, 1 (1988).

\bibitem{Royer}
A. Royer, J. Math. Phys. {\bf 24}, 897 (1983).

\bibitem{Laloe1995}
P. Gr\"uter, F. Lalo\"e, J. Physique I {\bf 5}, 181 (1995).

\bibitem{BedaqueRupak2003}
P.F. Bedaque, G. Rupak, Phys. Rev. B {\bf 67}, 174513 (2003).

\bibitem{Ouvry}
A. Comtet, Y. Georgelin, S. Ouvry, J. Phys. A {\bf 22}, 3917 (1989).

\bibitem{Cabe}
J. McCabe, S. Ouvry, Phys. Lett. B {\bf 260}, 113 (1990).

\bibitem{Law}
J. Law, Akira Suzuki, R.K. Bhaduri, 
Phys. Rev. {\bf A} 46, 4693 (1992).

\bibitem{Wilkens}
T. Busch, B.G. Englert, K. Rzazewski, M. Wilkens,
Found. Phys. {\bf 28}, 549 (1998).

\bibitem{WernerSym}
F. Werner, Y. Castin, 
Phys. Rev. A {\bf 74}, 053604 (2006).

\bibitem{GasaneoBirse}
G. Gasaneo, J.H. Macek, J. Phys. B {\bf 35}, 2239 (2002);
M. Birse, J. Phys. A {\bf 39}, L49 (2006).

\bibitem{Tignone}
Y. Castin, E. Tignone, Phys. Rev. A {\bf 84}, 062704 (2011).

\bibitem{Petrov_bosons}
D.S. Petrov, Phys. Rev. Lett. {\bf 93}, 143201 (2004).

\bibitem{Gogolin}
A.O. Gogolin, C. Mora, R. Egger, Phys. Rev. Lett. {\bf 100}, 140404 (2008).

\bibitem{Pricoupenko2010} 
L. Pricoupenko, Phys. Rev. A {\bf 82}, 043633 (2010).

\bibitem{Zwerger}
R. Schmidt, S.P. Rath, W. Zwerger,
Eur. Phys. J. B {\bf 85}, 386 (2012).

\bibitem{WernerThese}
F. Werner, PhD thesis, University Paris 6 (2008),
\url{http://tel.archives-ouvertes.fr/tel-00285587}

\bibitem{Larkin}
A.I. Larkin, 
Sov. Phys. JETP {\bf 11}, 1363 (1960). 

\bibitem{Gradstein}
I.S. Gradshteyn, I. M. Ryzhik, in \emph{Tables of Integrals, Series, and Products}, fifth edition,
edited by A. Jeffrey (Academic Press, San Diego, 1994).

\end{thebibliography}
\end{document}